\documentclass[12pt,preprint]{aastex} %{ctexart}
\usepackage{graphicx}
\usepackage{epstopdf}

\shorttitle{Flare of Sgr A*}
\shortauthors{Huang et al.}

\begin{document}

%% LaTeX will automatically break titles if they run longer than
%% one line. However, you may use \\ to force a line break if
%% you desire.

\title{CONSTRAINING THE FLARING REGION OF SAGITTARIUS A* BY 1.3MM VLBI MEASUREMENTS}

%% Use \author, \affil, and the \and command to format
%% author and affiliation information.
%% Note that \email has replaced the old \authoremail command
%% from AASTeX v4.0. You can use \email to mark an email address
%% anywhere in the paper, not just in the front matter.
%% As in the title, use \\ to force line breaks.

%\author{Lei Huang}

\author{Lei Huang\altaffilmark{1,2}, Zhi-Qiang Shen\altaffilmark{1,2}, Feng Gao\altaffilmark{1,3}}

\altaffiltext{1}{Key Laboratory for Research in Galaxies and
Cosmology, Shanghai Astronomical Observatory, Chinese Academy of
Sciences, Shanghai 200030, China; muduri@shao.ac.cn; zshen@shao.ac.cn; fgao@shao.ac.cn}

\altaffiltext{2}{Key Laboratory of Radio Astronomy, Chinese Academy of Sciences, China}

\altaffiltext{3}{Graduate School of the Chinese Academy of Sciences, Beijing 100039, China}

%% Mark off your abstract in the "abstract'' environment. In the manuscript
%% style, abstract will output a Received/Accepted line after the
%% title and affiliation information. No date will appear since the author
%% does not have this information. The dates will be filled in by the
%% editorial office after submission.

\begin{abstract}

We use a model of an accretion flow coupled with an emergent flare to interpret the latest 1.3mm VLBI measurements for Sagittarius A*. The visibility data constrained the distances from the flare center to the black hole center as $d_{\rm EW}\lesssim20{\rm R_g}$ and $d_{\rm NS}\lesssim80{\rm R_g}$ in the East-West and North-South directions, respectively. If interpreted by the hot-spot model, the flare was preferred to pass in front of the black hole at a radius much larger than $d_{\rm EW}$. If interpreted by the episodic jet launched from a nearly edge-on hot accretion flow, the flare was preferred to be ejected with $\theta_{\rm j}\gtrsim40^\circ$ off the black hole rotating axis.  This method can be generalized to help us understand future sub-millimeter VLBI observations, and study the millimeter/sub-millimeter variabilities in the vicinity of the Galactic Center supermassive black hole.

\end{abstract}


\section{INTRODUCTION}
\label{intro}

Sagittarius A* (Sgr A*) is the best black hole candidate that can be studied by Very Long Baseline Interferometry (VLBI).   \citet{Doeleman08} reported 1.3mm VLBI detections of Sgr A* on two baselines, namely ARO/SMT (Arizona) - CARMA (California) and ARO/SMT - JCMT (Hawaii).  These data were fitted by a circular Gaussian component with a full-width of half maximum (FWHM) of $\sim43\mu{\rm as}$, reaching the event-horizon scale of the Galactic Center supermassive black hole.  Several groups soon interpreted these data independently by an accretion flow around a rotating black hole with a high inclination angle \citep{Broderick09, Huang09a, Huang09b, Dexter09}.

Recently, \citet{Fish11} reported new measurements on all the three baselines of the ARO/SMT-CARMA-JCMT array over three nights: 2009 April 5-7 (Days 95-97).  The data on Day 95 and Day 96 show high consistency with the data in 2007 April by \citet{Doeleman08}, fitted by a circular Gaussian component with FWHMs of $\sim41\mu{\rm as}$ and $\sim44\mu{\rm as}$, respectively. But the correlated flux density on all the baselines increased on Day 97.  The Gaussian model with a similar size but higher flux density is needed to fit these data.  However, they pointed out that some model parameters must be changed for Day 97 in order to fit the data on all the three days by a ring model.   This implies that an additional variable component might exist together with the accretion flow associated with a black hole shadow structure.

In this Letter, we introduce a method in Section 2 to constrain the position of the flaring component, as required to interpret the Day 97 data.   In Section 3, we hold general discussions on the resolution of the baselines, challenges to the flare models, and suggestions for future observations.  Summary is provided in Section 4.   Here we adopt black hole mass $M_{\rm BH}=4.1\times10^6M_\odot$ \citep{Ghez08} and distance of the Galactic Center $d_{\rm GC}=8{\rm kpc}$. Thus 1 milli-arcsecond (mas) angular size corresponds to a linear size at the Galactic Center of $200{\rm R_g}$, where ${\rm R_g}=GM_{\rm BH}/c^2$ is the gravitational radius.

\newpage

\section{CONSTRAINING THE POSITION OF FLARING COMPONENT}
\label{position}

\begin{figure}[ht]
\vspace{-0mm}
\begin{center}
\includegraphics[width=12cm]{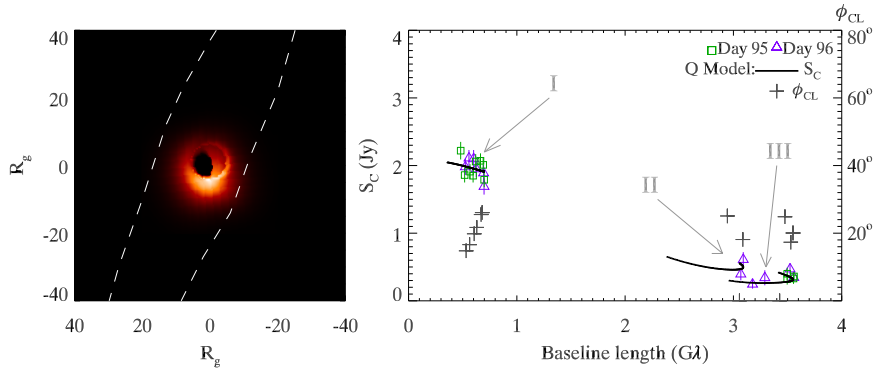}
\vspace{-5mm}\caption{ \textit{Left}: Non-scattered image of accretion flow associated with a black hole shadow structure (Q Model).
\textit{Right}: Corresponding visibility amplitudes ${\rm S_C}$ (black lines) and closure phases $\phi_{\rm CL}$ (plus signs) of baselines I, II and III.  Visibility data are shown in squares for Day 95 and in triangles for Day 96.
\label{meas9596}}
\end{center}
\end{figure}

Hereafter, for simplicity, we denote the three baselines from short to long, namely ARO/SMT-CARMA, CARMA-JCMT, and ARO/SMT-JCMT by baselines I, II, and III, respectively.  We apply all the data on Days 95-97 \citep{Fish11} and take averages if there are more than one data points measured on the same baseline coordinates $(u,v)$ on the UV plane.  We interpret the measurements on Day 95 and Day 96 as contributions by an accretion flow in its quiescent state.   The magneto-rotational-instability (MRI) dominated relativistic accretion flow \citep{Huang09b}, named as Q Model, with black hole spin $a=0.5$, inclination angle $i=60^\circ$, and position angle $\Theta=100^\circ$, is adopted to account for the emission in the quiescent state.  The black hole spin and orientation are typical in our earlier work, which are also in agreement with the estimates made by other groups \citep{Dexter10, Shch10, Broderick11}.

We simulated the image (shown in the left panel of Fig.\ref{meas9596}) of the accretion flow by ray-tracing with polarized relativistic radiative transfer \citep{Shch11, Huang11}.  We further plot the visibility data of this theoretical image in right panel of Fig.\ref{meas9596}, with inter-stellar scattering considered \citep{Shen05}. The total flux density ${\rm S_{Q0}}=2.1{\rm Jy}$.  The amplitude of visibility, i.e. the correlated flux density ${\rm S_C}$ shown in black lines fit the data on Day 95 with reduced chi-squares $\chi^2_{\rm dof}=1.06$ and those on Day 96 with $\chi^2_{\rm dof}=1.07$, here ${\rm dof=12}$ on both days.  The closure phases $\phi_{\rm CL}$ of baselines I, II, and III averaged over 10 minutes, as shown in plus signs, is in the range $({0^\circ},{25^\circ})$, which is also consistent with the observations at $0^\circ\pm40^\circ$ \citep{Fish11}.

\newpage

\begin{figure}[ht]
\vspace{-0mm}
\begin{center}
\vspace{-2.1cm} \includegraphics[width=12cm]{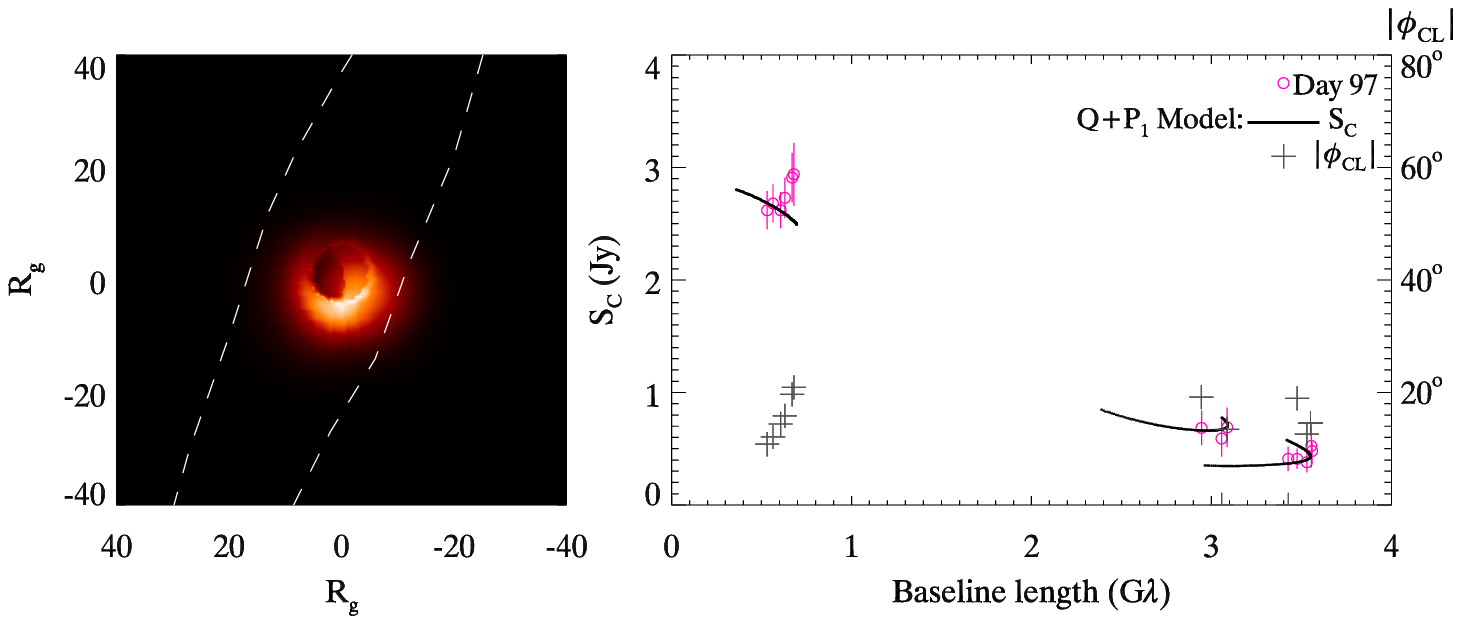} \\
\vspace{-1.6cm} \includegraphics[width=12cm]{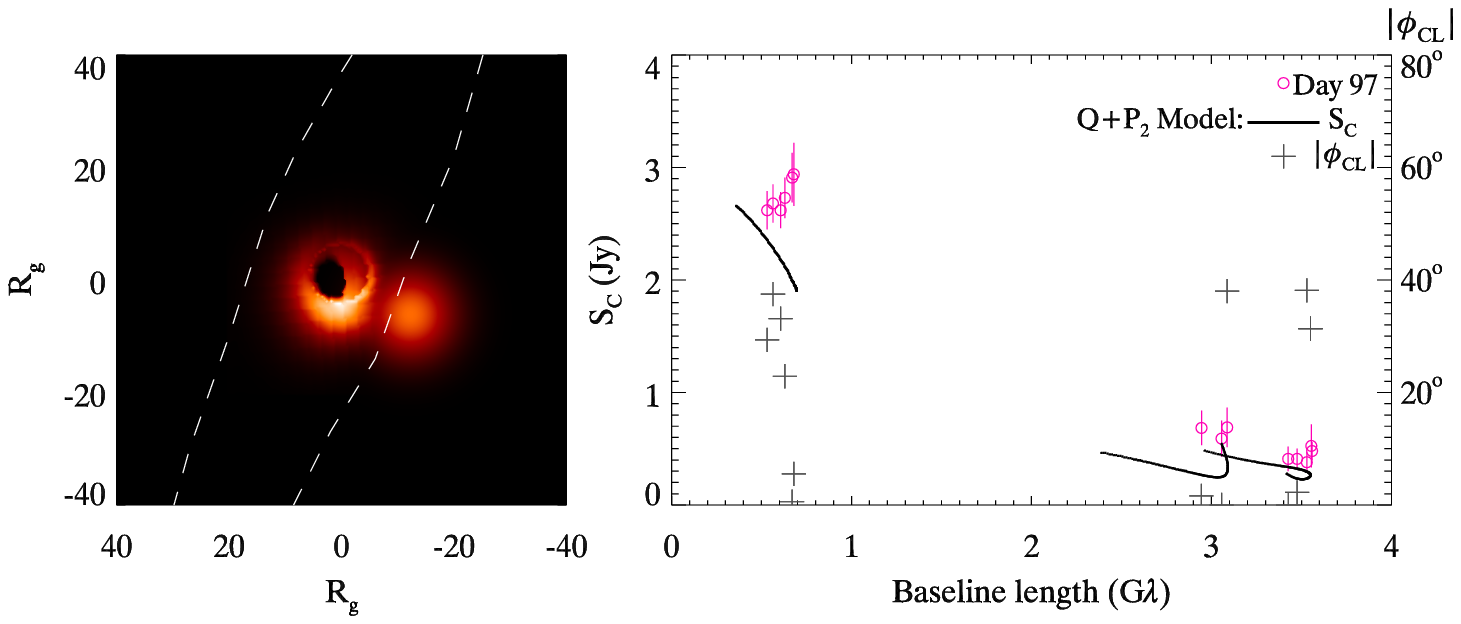} \\
\vspace{-1.6cm} \includegraphics[width=12cm]{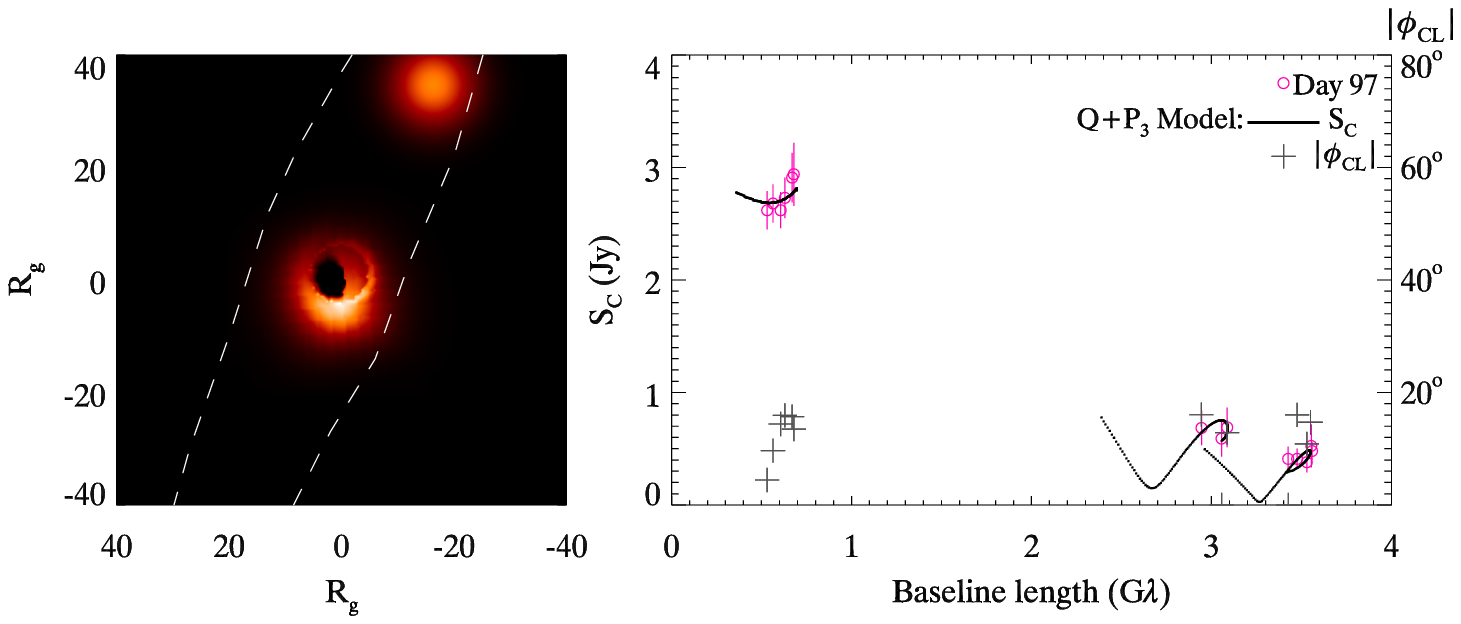}
\vspace{- 5mm}\caption{ \textit{Left Panels}: Non-scattered images of accretion flow coupled with a flare with different central shifts. Boundaries of preferred positions of flare are shown in dashed white lines.  \textit{Right Panels}: Visibility amplitudes and closure phases corresponding to left images. Absolutes of $\phi_{\rm CL}$ are taken for a concise layout.  Visibility data are shown in circles for Day 97. Cases shown in the \textit{top} and \textit{bottom} panels are consistent with the observations, while case in the \textit{middle} panel is not.  See text for more details.
\label{meas97eg}}
\end{center}
\end{figure}

We ascribe the increase of visibility on Day 97 relative to Days 95-96 to the emergence of a flaring component.  For simplicity, we assume that the brightness distribution of this flare can be represented by a circular Gaussian function ${\mathcal B}_{\rm P}({\rm S_{P0}},{\rm D_P},x_{\rm P},y_{\rm P})$ with four parameters, namely ${\rm S_{P0}}$ (total flux density), ${\rm D_p}$ (FWHM), $x_{\rm P}$ (shift of flare center) in abscissa, and $y_{\rm P}$ (shift in ordinate).   The brightness distribution of the flaring state of the accretion flow,  ${\mathcal B}_{\rm F}$, is simply assumed to be the superposition of the brightness distribution of the quiescent state,  ${\mathcal B}_{\rm Q}$, and that of the flare,  ${\mathcal B}_{\rm P}$, i.e. ${\mathcal B}_{\rm F}={\mathcal B}_{\rm Q}+{\mathcal B}_{\rm P}$.  The visibility distribution with inter-stellar scattering considered, ${\mathcal V}_{\rm F}={\mathcal F}\left\{{\mathcal B}_{\rm F, scat}\right\}$, is also the sum of ${\mathcal V}_{\rm Q}={\mathcal F}\left\{{\mathcal B}_{\rm Q, scat}\right\}$ and ${\mathcal V}_{\rm P}={\mathcal F}\left\{{\mathcal B}_{\rm P, scat}\right\}$, according to the property of linearity of Fourier Transform ${\mathcal F}$. By the property of translation of Fourier Transform, we further obtain
\begin{eqnarray}
	{\mathcal V}_{\rm F}&=&{\mathcal V}_{\rm Q}+{\mathcal V}_{\rm P0}\nonumber\\
	{\mathcal V}_{\rm P0}&=&{\mathcal F}\left\{{\mathcal B}_{\rm P, scat}({\rm S_{P0}},{\rm D_P},0,0)\right\}{\rm e}^{-2\pi{\rm i}(x_{\rm P}{\cdot}u+y_{\rm P}{\cdot}v)},
\end{eqnarray}
where $u$ and $v$ are the baseline coordinates on the UV plane.

We make a search over wide ranges of the parameters, ${\rm S_{P0}}$, ${\rm D_P}$, $x_{\rm P}$, and $y_{\rm P}$, and fit the data on Day 97 to calculate the reduced chi-square $\chi^2_{\rm dof}$ with ${\rm dof}=N_{\rm data}-N_{\rm param}=14-4=10$.  We choose sets of parameters that yield $\chi^2_{\rm dof}=1\pm1$ by preference.  Thus, we obtain ${\rm S_{P0}}\sim0.95\pm0.55$Jy and  ${\rm D_P}\sim6.2\pm6.2{\rm R_g}$.  We estimate the observational angular size of the flare as $\theta_{\rm obs}=38^{+28}_{-17}{\rm mas}$ with the formula $\theta_{\rm obs}^2=\left({\rm D_p} \cdot\theta_{\rm Rg}\right)^2+\theta_{\rm scat}^2$, where $\theta_{\rm Rg}\approx0.005{\rm mas}$ is the angular size of gravitational radius and $\theta_{\rm scat}\approx0.021{\rm mas}$ is the extrapolated angular size of scattering screen in the EW direction \citep{Shen05}.  The corresponding brightness temperature of the flare is $T_b=1.22\times10^{12}{\rm (S_0 /Jy)}{(\nu/{\rm GHz})}^{-2}{(\theta_{\rm obs}/{\rm mas})}^{-2}\sim1.5^{+0.5}_{-0.7}\times10^{10}{\rm K}$.  This value is consistent with the estimation for a plasmon adopted to interpret the variability of Sgr A* on 3mm wavelength observed by ATCA \citep{Li09}.  We show the boundaries of the preferred positions in dashed white lines in the left panels of Fig.\ref{meas97eg}, i.e. all the preferred pairs of $(x_{\rm P},y_{\rm P})$ located in the region surrounded by the boundaries.  This region is narrow along the direction of $\sim65^\circ$ east-of-north, hereafter denoted as EW direction, and elongated along the direction of $\sim25^\circ$ west-of-north, hereafter as NS direction.  The region has an asymmetrical shape that slightly varies with the parameters of the Q model.  Generally speaking, the preferred distances from the flare center to the black hole center are $d_{\rm EW}\lesssim20{\rm R_g}$ and $d_{\rm NS}\lesssim80{\rm R_g}$, along the EW and NS directions, respectively.

The data on Day 97 impose a strict constraint on $d_{\rm EW}$ but a relatively relaxed one on $d_{\rm NS}$.  This is mainly because the measurements on the tracks of the two long baselines II and III are roughly aligned in the EW direction.  These baselines are longer than $3{\rm G}\lambda$, giving a high angular resolution of $85 \mu{\rm as}$, or $17 {\rm R_g}$, in the EW direction.  A flaring component emerging at  $d_{\rm EW}\gtrsim17{\rm R_g}$ would cause a null/valley point in the visibility profile at a baseline length $\lesssim3{\rm G}\lambda$, so that the amplitudes of visibility on baseline II might be lower than those on baseline III.  However, the amplitudes actually have a monotonically decreasing profile from baseline II to baseline III, which implies there is no structural variability with scales larger than $17{\rm R_g}$ in the EW direction.   Moreover, the track of baseline I can also detect the separation of the two components in the EW direction sensitively,  although its precise direction is $\sim70^\circ$ west-of-north, which is $\sim45^\circ$ off those of baselines II and III.    On the contrary, $d_{\rm NS}$ cannot be well-constrained since there are no long enough projected baselines available in the NS direction.

In the left panels of Fig.\ref{meas97eg}, we show three examples of the flaring state, each consisting of the accretion flow overlaid by a different flare.  ${\rm P_1}$, shown in the top-left panel, with ${\rm S_{P10}}=0.8$Jy, ${\rm D_{P1}}=6.6 {\rm R_g}$, and $(x_{\rm {P1}},y_{\rm {P1}})/{\rm R_g}=(-4.08,-2.04)$, is located within the preferred region.  The visibility corresponding to ${\rm Q+P_1}$ is shown in the top-right panel, fitting the data on Day 97 by $\chi^2_{\rm dof}=1.13$.  The closure phase $\phi_{\rm CL}$ is within the range $(-20^\circ,20^\circ)$,  indicating high symmetry of the total image.  ${\rm P_2}$, shown in the middle-left panel, with ${\rm S_{P20}}=0.8$Jy, ${\rm D_{P2}}=6.6{\rm R_g}$, and $(x_{\rm {P2}},y_{\rm {P2}})/{\rm R_g}=(-12.24,-6.12)$, is outside the preferred region.  The visibility fits the data by $\chi^2_{\rm dof}=8.27$.  As shown in the middle-right panel, a flare slightly beyond the boundaries in the EW direction can decrease ${\rm S_C}$ a lot on baseline I and cause a valley structure on baseline II.  $\phi_{\rm CL}$ is in the range $(-40^\circ,40^\circ)$, deviating greatly from zero, and inferring that the symmetry of the total image in the EW direction is somehow broken.  ${\rm P_3}$, shown in the bottom-left panel, with ${\rm S_{P30}}=0.95$Jy, ${\rm D_{P3}}=6.2{\rm R_g}$, and $(x_{\rm {P3}},y_{\rm {P3}})/{\rm R_g}=(-16.34,34.68)$, is within the preferred region.  This is the case with minimal $\chi^2_{\rm dof}=0.25$, i.e. the best-fit.  As shown in the bottom-right panel, it even reproduces the $\sim10\%$ increase in ${\rm S_C}$ related to baseline length on baseline I.  However, considering the uncertainties of the data and the simplicity of the static models adopted here, we think it may have over-fitted the data, especially on baseline I.

\newpage

\section{DISCUSSION}
\label{discuss}

We introduce a simple but useful method to understand the time-variable emission of Sgr A* detected by 1.3mm VLBI \citep{Fish11}.  We interpret the data on Days 95-96 by an accretion flow in its quiescent state, and the increase of correlated flux density on all the three baselines on Day 97 by the emergence of an extra flare, modeled by a circular Gaussian spot.  The visibility measurements impose a strict constraint on the distance from the flare center to the black hole center in the EW direction ($65^\circ$ east-of-north) as $d_{\rm EW}\lesssim20{\rm R_g}$. They also place another relatively relaxed constraint on the distance in the NS direction ($25^\circ$ west-of-north) as $d_{\rm NS}\lesssim80{\rm R_g}$. In the rest of this Section, general discussions on various aspects will follow.

\subsection{Resolution of Baselines}

The longest baseline length on the tracks of baseline III (ARO/SMT-JCMT) is $\sim3.55{\rm G}\lambda$, corresponding to the highest resolution of $\theta_{\rm beam} \sim70\mu{\rm as}$.  The apparent size of the flaring component assumed in this Letter is $\lesssim66\mu{\rm as}$, slightly super-resolved by the current baseline.  Here, we follow the criterion provided in \citet{Shen97} to judge the degree of resolution.

$\theta_{\rm LIM}$, the limit of source size that can be resolved, is defined as
\begin{eqnarray}
	&\quad&\theta_{\rm LIM}\quad=\quad\left[\theta_{\rm LIM}({\rm Statistical})^4+\theta_{\rm LIM}({\rm Systematic})^4\right]^{1/4},\nonumber 
\end{eqnarray}
where
\begin{eqnarray}
	&\quad&\theta_{\rm LIM}({\rm Statistical})\quad=\quad\frac{0.53}{\sqrt{\rm SNR}}\cdot\theta_{\rm beam}\nonumber\\
	&\quad&\theta_{\rm LIM}({\rm Systematic})\quad=\quad\frac{0.53}{\sqrt{|F_\nu/\Delta{F_\nu}}|}\cdot\theta_{\rm beam}.
\end{eqnarray}
With a mean signal-to-noise ratio (SNR) of 5.5 on the ARO/SMT-JCMT baseline \citep{Doeleman08} and $|\Delta{F_\nu}/F_\nu|$ being the sum of the uncertainties of the flux density measurements $\sim30\%$, this limit is estimated to be $\theta_{\rm LIM}\sim20\mu{\rm as}$.  Thus, we think the flare was resolved in an optimistic manner.

\newpage

\subsection{Challenges To The Flare Models}
\label{origin}

The origin of flares in Sgr A* is still controversial.  Various models can be put into two general classes, namely the whole change and the transient structure.  The whole change means enhancement in flux density in the whole emission region, caused by changes in physical quantities of the accretion flow, e.g. increase in heating coefficient or creation of power-law electrons \citep[e.g.][]{Yuan03, Liu04}.  We can fit the Day 97 data by the Q Model itself with $\chi^2_{\rm dof}=1.07$, ${\rm dof}=N_{\rm data}=14$, if the total flux density is scaled up to $3$Jy.  This implies those models of whole change are plausible for the Day 97 data, i.e. no extra component is required for data interpretation.  This is consistent with the result of $d_{\rm EW}\lesssim20{\rm R_g}$, indicating that the flare happened very close to, or even inside the emission region of accretion flow. 

We are also interested in challenging models of the transient structure, which means an extra component emerged to contribute to structural variability.  Generally speaking, the Day 97 data do not preclude any flare model if the transient structure is located within the preferred flaring region.

A popular model used to interpret short-time variability of Sgr A* in millimeter and near-infrared bands is a hot-spot orbiting in Keplerian angular velocity \citep[e.g.][]{Broderick05, Dovciak08}.  Such a hot-spot varying in hourly timescale is predicted to be detectable by sub-mm VLBI \citep{Doeleman09}.   We assume that the flare on Day 97 was caused by a hot-spot orbiting at ${\rm r_\parallel}$,  the projective radius from the black hole center.  The normal direction of the orbital plane coincides with the projective rotating axis of the black hole.  According to earlier work, the position angle of Sgr A* is preferred to be in the range $(90^\circ,180^\circ)$ or $(-90^\circ,0^\circ)$, rather than in the range $(0^\circ,90^\circ)$ or $(-180^\circ,-90^\circ)$. We then obtain ${\rm r_\parallel<40R_g}$, as calculated from the constraints on the flaring region shown in the above Section.  The period of the Keplerian orbit is constrained as ${\rm T_{Kep}(r=r_\parallel ) < 8hr}$, which is insensitive to black hole spin.  The observational duration $\rm T_{obs}$ was $\sim2{\rm hr}$ on Day 97,  i.e., ${\rm T_{obs}>0.25{\cdot}T_{Kep}(r=r_\parallel)}$, implying that an apparent light-curve should be detected with a nearly edge-on disk assumed.  However, this is inconsistent with the observed sustaining high flux density.  Therefore, we would either exclude this model, or explain this inconsistency by two arguments.  One is that the measurements of the total flux density had rather large errors. The other is that the radius was shortened by projection effect, i.e. the hot-spot might be passing in front of the black hole at a radius ${\rm r_\perp>40R_g}$ with a component aligned with the line-of-sight.

We further generalize this model into a hot-spot moving with sub-Keplerian angular velocity and radial velocity comparable to the light speed, the same as the velocity of the background transonic accretion flow.  In this case, the hot-spot would quickly fall into the event horizon of the black hole and disappear.  We integrate the curve of radial velocity shown in \citet{Huang09b} to calculate typical timescale of falling at specific starting radii ${\rm r_\parallel<40R_g}$.  The lifetime of the hot-spot is constrained as ${\rm T_{fall}(r=r_\parallel)<1hr}$,  i.e. ${\rm T_{obs}>2{\cdot}T_{fall}}$.  Similar to the Keplerian-rotating model, we would either exclude this model or interpret it as a special case in which the hot-spot started at a larger radius in front of the black hole with ${\rm r_\perp>60R_g}$, and plunged into the event horizon.

Furthermore, we consider a different model of episodic jets proposed by \citet{Yuan09}, in which the flare is contributed by an episodic ejection of plasmoid from a hot accretion flow.  As they calculated,  a plasmoid with initial location of $\sim10{\rm R_g}$ to Sgr A*  can accelerate from rest to $\sim0.8c$ in 35 min.  This predicts that during the 2h observation,  the plasmoid can travel $\sim250\rm R_g$ from the black hole center.  To take into account both the constraint on $d_{\rm NS}$ and the projection effect by a disk with $i\gtrsim60^\circ$,  the plasmoid must be ejected in a direction with $\theta_{\rm j}\gtrsim40^\circ$ off the rotating axis of the black hole.

We wish to make a comment on a model of tidal disruption of asteroid explored by \citet{Zubovas11}, in which the flare is related to asteroids rather than properties of hot accretion flow.  They predicted a small size for a flaring region $\lesssim1{\rm AU}$ or $\lesssim25{\rm R_g}$, which is roughly included in our constraints with boundaries of $d_{\rm EW}$ and $d_{\rm NS}$.  However, the timescale of an asteroid in a parabolic orbit around Sgr A* is estimated to be $\lesssim1.5{\rm hr}$, which is unfavored by the Day 97 flare.

\newpage

\subsection{Visibility Prediction For An Array of Five Stations}

\begin{figure}[ht]
\vspace{-0mm}
\begin{center}
\vspace{-1cm} \includegraphics[width=12cm]{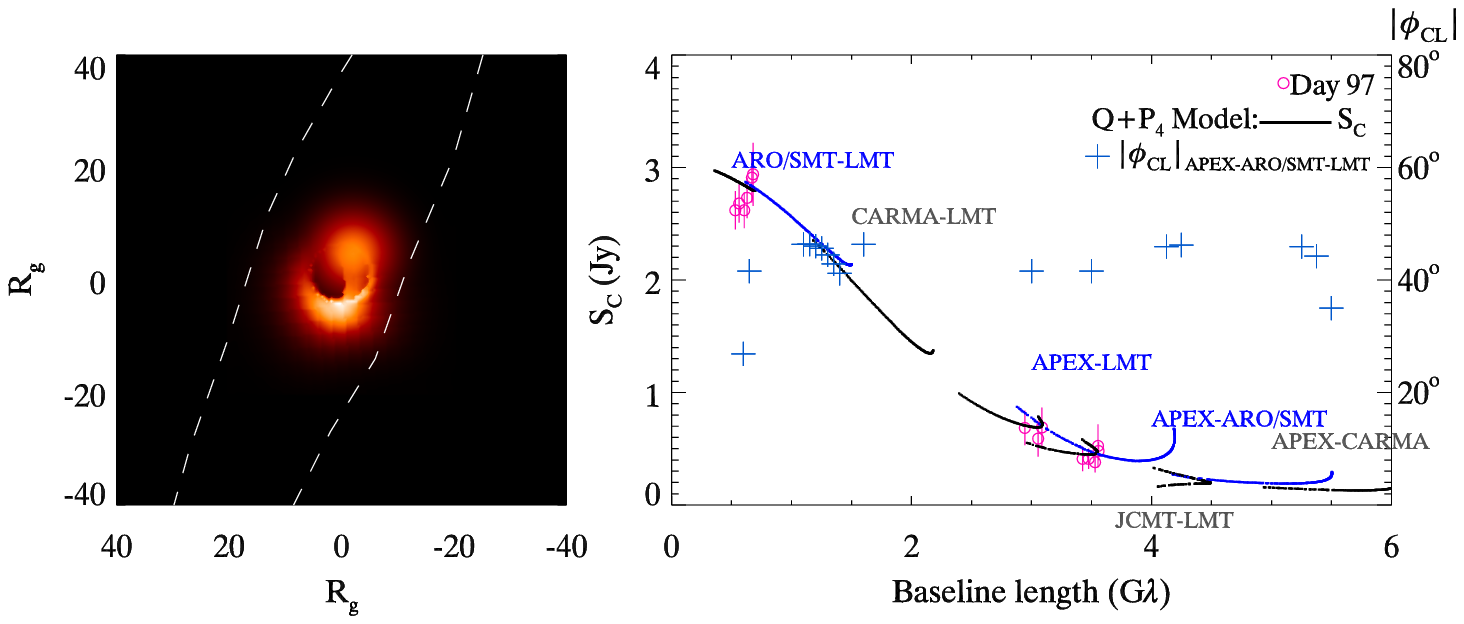} \\
\vspace{-1.2cm} \includegraphics[width=12cm]{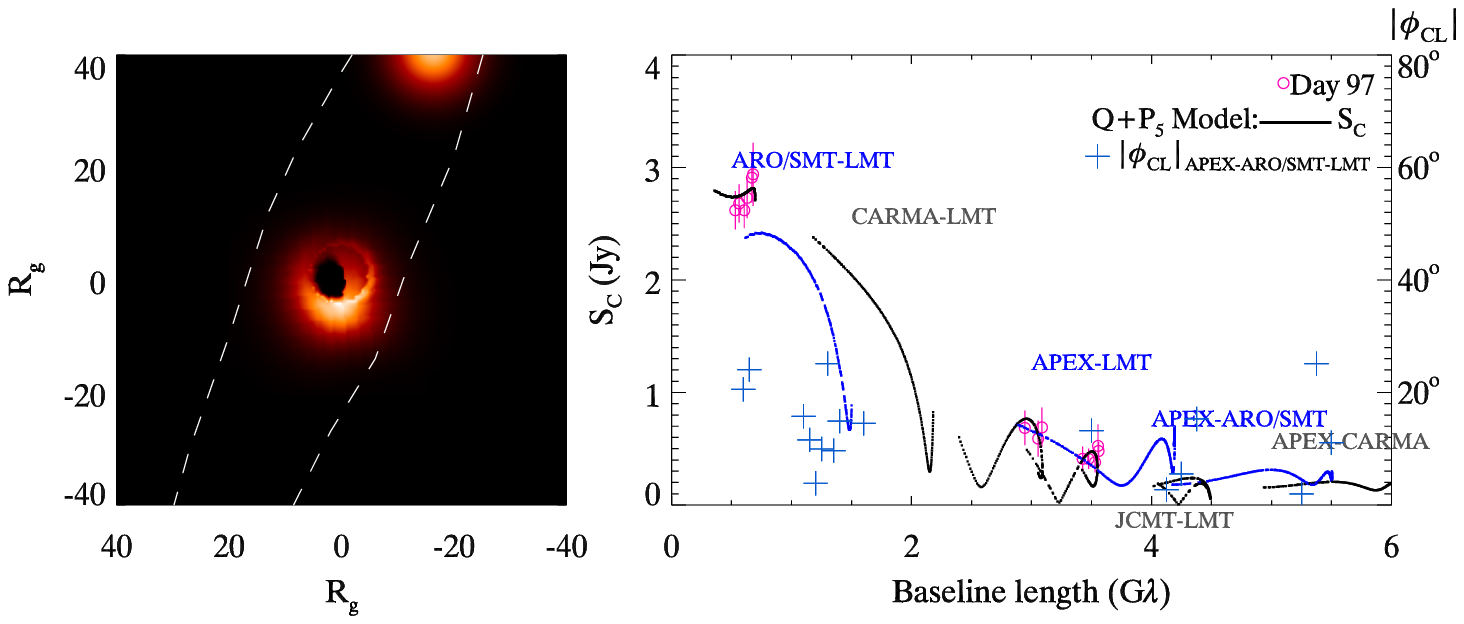} \\
\vspace{-5mm}\caption{ \textit{Left Panels}: Non-scattered images of accretion flow coupled with a flare with different central shifts. Boundaries of preferred positions of the flare are shown in dashed white lines.  \textit{Right Panels}: Visibility amplitudes and closure phases (APEX-ARO/SMT-LMT, in plus signs) corresponding to the left images. See text for more details.
\label{meas975sta}}
\end{center}
\end{figure}

The flaring region of Sgr A* can be constrained better if suitable baselines in the NS direction are available. Theoretically, the separation of $\Delta S\approx d_{\rm NS}\lesssim80{\rm R_g}$ between two components, the quiescent accretion flow and the flare, will cause first null/valley point in the visibility profile at $L_0\gtrsim0.25{\rm G}\lambda$ in the NS direction.  We include two additional stations, APEX in Chile and LMT in Mexico, which are promising for contributing to sub-mm VLBI observations in the near future.  The tracks of seven additional baselines, from short to long, ARO/SMT-LMT, CARMA-LMT, APEX-LMT, JCMT-LMT, APEX-ARO/SMT, APEX-CARMA, and APEX-JCMT,  cover from $0.5{\rm G}\lambda$ to $7{\rm G}\lambda$ in length close to the NS direction.  In an optimistic view, the separation of $\Delta S\approx d_{\rm NS}\lesssim40{\rm R_g}$ could be detected if all the baselines work.  In the left panels of Fig.\ref{meas975sta}, we choose two examples of flares, ${\rm P_4}$ and ${\rm P_5}$, both with ${\rm S_p=0.95Jy}$ and ${\rm D_p=6.2R_g}$, but with $(x_{\rm {P4}},y_{\rm {P4}})/{\rm R_g}=(-2.04,4.9)$ and $(x_{\rm {P5}},y_{\rm {P5}})/{\rm R_g}=(-16.32, 40.8)$, respectively.  The corresponding visibilities are shown in the right panels.  The image of the Q Model and ${\rm P_4}$, with $\Delta S\approx5{\rm R_g}$, causes the first valley point at $\sim4{\rm G}\lambda$ on the baseline APEX-LMT.   The image of the Q Model and ${\rm P_5}$, with $\Delta S\approx40{\rm R_g}$, causes the first valley point at $\sim0.5{\rm G}\lambda$ on the baseline ARO/SMT-LMT, which appears as a drop in amplitude between the baselines ARO/SMT-CARMA and ARO/SMT-LMT.  At the same time, closure phases by a station group including APEX and LMT, also imply the change to symmetry caused by the flare. For example, $|\phi_{\rm CL}|$ by the group of APEX - ARO/SMT - LMT are shown in blue plus signs in the right panels of Fig.\ref{meas975sta}.

\section{SUMMARY}

The detection of time-variable emission of Sagittarius A* by 1.3mm VLBI indicates instability on event-horizon scales.  Modeling the flare by a circular Gaussian spot, the data constrained the spot size as $\rm D_P\sim6.2\pm6.2R_g$ and deviation to the black hole center in the East-West direction ($65^\circ$ east-of-north) as $d_{\rm EW}\lesssim20{\rm R_g}$, both being comparable to the size of the black hole shadow ($\sim\rm10R_g$).  Alternatively, we can interpret this flaring activity as enhancement in whole emission region, and we would prefer such an interpretation to models of transient structures that require an extra component.   This is mainly due to the sustaining high flux density in long observational duration, $\rm T_{obs}\sim2hr$.  The duration is more than one-fourth of the orbiting period if modeled by a hot-spot in Keplerian rotation  at a preferable orbiting radius $\lesssim \rm40R_g$. It is also more than twice the falling timescale if modeled by a hot-spot with sub-Keplerian angular velocity and high radial velocity, and with a preferable initial radius $\lesssim\rm 40R_g$.  A hot-spot located at $d_{\rm EW}$ cannot maintain high flux density during the observational duration, unless it passed in front of the black hole at a radius much larger than $d_{\rm EW}$ so that the observed distance is shortened by projection effect. If interpreted by an episodic jet, the ejecting plasmoid cannot be confined within the preferred flaring region with an assumed edge-on accretion flow, unless the angle between the ejecting direction and the black hole rotating axis is greater than $40^\circ$. This method of visibility analysis can be generalized for future sub-millimeter VLBI measurements. We would have a better understanding of the nature of variability in the black hole vicinity of the Galactic Center if $d_{\rm NS}$, deviation of the flare to the black hole center in the North-South direction ($25^\circ$ west-of-north), could be constrained more precisely with new stations included.

\acknowledgments

This work was supported in part by the National Natural Science
Foundation of China (grants 10625314, 10821302 and 11173046),
the National Key Basic Research Development
Program of China (No. 2012CB821800), and the CAS/SAFEA International
Partnership Program for Creative Research Teams.

%\clearpage

\vspace{5cm}

\end{document}